\documentclass{article}

\usepackage{amsmath}
\usepackage{amssymb}
\usepackage{cite}

\def\d{{\mathrm{d}}}
\def\S{{\mathcal{S}}}

\begin{document}
\begin{titlepage}

\begin{center}
{\Large\textbf{The Fokker-Planck equation, and stationary
densities}}

\vskip 2 cm

{Amir~Aghamohammadi{\footnote {mohamadi@alzahra.ac.ir}} \&
Mohammad~Khorrami{\footnote {mamwad@mailaps.org}}} \vskip 5 mm

\textit{  Department of Physics, Alzahra University,
             Tehran 1993891167, Iran. }
\end{center}

\begin{abstract}
\noindent The most general local Markovian stochastic model is
investigated, for which it is known that the evolution equation is
the Fokker-Planck equation. Special cases are investigated where
uncorrelated initial states remain uncorrelated. Finally,
stochastic one-dimensional fields with local interactions are
studied that have kink-solutions.

\end{abstract}
\begin{center} {\textbf{PACS numbers:}} 05.40.-a, 02.50.Ga

\end{center}
\end{titlepage}
\section{Introduction}
The Fokker-Planck equation  was first used to describe the
statistics of the Brownian motion of a particle in a fluid. This
equation describes the time evolution of the probability density
function of the position or the velocity of a Brownian particle.
It is a differential equation, second order in the spatial
variables and first order in time. There are also some studies on
generalizations of the Fokker-Planck equation involving spatial
derivatives of order higher than two
\cite{Kamp,Leb1,Leb2,Leb3,Ply}. Although these higher order
equations are linear in the distribution function, these are
sometimes referred to as nonlinear Fokker-Planck equations. There
are also other generalizations of the Fokker-Planck equation
\cite{Tara,Chav}. In \cite{Paw}, it has been shown that any
truncation of the generalized Fokker-Planck equation (the
Kramers-Moyal expansion) is inconsistent, unless only terms of up
to second order derivatives are kept.

Reaction-diffusion systems is a well-studied area. People have
studied reaction-diffusion systems using analytical techniques,
approximation methods, and simulations. A large fraction of exact
results belong to low-dimensional (specially one-dimensional)
systems, as solving low-dimensional systems should in principle be
easier. Despite their simplicity, these systems exhibit a rich and
rather nontrivial dynamical and stationary behavior. Studies on
the models far from equilibrium have shown that there is a
remarkably rich variety of critical phenomena \cite{ScR}. There
are some exact results involving kinks in one-dimensional
reaction-diffusion models, as well as simulations and numeric
results \cite{J1}, and also mean field results \cite{PS2}.
Formation of localized shocks in one-dimensional driven diffusive
systems with spatially homogeneous creation and annihilation of
particles has been studied in \cite{PRW}. Recently, in \cite{KJS},
the families of models with travelling wave solutions on a finite
lattice have been presented. Shocks have been studied at both the
macroscopic and the microscopic levels and there are efforts to
understand how these macroscopic shocks originate from the
microscopic dynamics \cite{PS2}. Hydrodynamic limits have also
been investigated. In a recent article both stationary and
dynamical single-kinks on a one-dimensional lattice have been
investigated \cite{AA}. It was done for both an infinite lattice
and a finite lattice with boundaries. Static and dynamical phase
transitions of these models have also been studied.

In such studies, one way is to investigate the system on a lattice
and then take the limit of the results in the continuum. Another
way is to write the equation in continuum from the beginning. In
the latter case, one should write the master equation in
continuum. This problem is addressed in this paper, and based on
the results simple examples are studied.

The scheme of the paper is the following. In section 2, the
Fokker-Planck equation and its generalizions are briefly reviewed,
where it is known that if the evolution equation of the
probability density is local in space, then the evolution equation
involves no more than second spatial derivatives \cite{Paw}. By
locality, it is meant that the time derivative of probability
density is expressible in terms of only a finite number of spatial
derivatives of that density. This means that the evolution
equation for the most general local stochastic Markovian models
with is the Fokker-Planck equation. In section 3, this equation is
generalized to the case of random fields. It is seen that there
are special local interactions for which uncorrelated initial
states remains uncorrelated. Finally, section 4 is devoted to some
simple examples of stocahstic fields, admitting kink solutions.

\section{The Fokker-Planck equation}
To fix notation, let's briefly review the Fokker-Plank equation.
Let $\mathbf{r}$ be a continuous ($D$-dimensional) random
variable, and $\psi(\mathbf{r},t)$ be the corresponding
probability density at the time $t$. One has
\begin{align}\label{fp01}
\psi(\mathbf{r},t)\geq& 0,\nonumber\\
\int\d^Dr\;\psi(\mathbf{r},t)=&1.
\end{align}
The evolution equation for this probability density is
\begin{equation}\label{fp02}
\psi(\mathbf{r}, t)= \int
\d^Dr'\;U(\mathbf{r},\mathbf{r}';t,t')\,\psi(\mathbf{r}',t'),
\end{equation}
where $U$ is the evolution operator with the following properties
\begin{align}\label{fp03}
U(\mathbf{r},\mathbf{r}';t,t')&\geq 0,\nonumber\\
\int\d^Dr\;U(\mathbf{r},\mathbf{r}';t,t')&=1,\nonumber\\
U(\mathbf{r},\mathbf{r}';t,t)&=\delta(\mathbf{r}-\mathbf{r}').
\end{align}
Differentiating (\ref{fp02}) with respect to $t$, one arrives at
\begin{equation}\label{fp04}
\frac{\partial\psi(\mathbf{r}, t)}{\partial t}= \int
\d^Dr'\;\frac{\partial U(\mathbf{r},\mathbf{r}';t,t')}{\partial
t}\,\psi(\mathbf{r}',t'),
\end{equation}
or
\begin{equation}\label{fp05}
\frac{\partial\psi(\mathbf{r}, t)}{\partial t}= \int
\d^Dr'\;H(\mathbf{r},\mathbf{r}';t)\,\psi(\mathbf{r}',t),
\end{equation}
where $H$ is defined through
\begin{equation}\label{fp06}
\frac{\partial U(\mathbf{r},\mathbf{r}';t,t')}{\partial t}=: \int
\d^Dr''\;H(\mathbf{r},\mathbf{r}'';t)\,U(\mathbf{r}'',\mathbf{r}';t,t'),
\end{equation}
or
\begin{equation}\label{fp07}
H(\mathbf{r},\mathbf{r}';t):=\left.\frac{\partial
U(\mathbf{r},\mathbf{r}';t,t')}{\partial t}\right|_{t'=t}.
\end{equation}
Using (\ref{fp03}), one obtains
\begin{align}\label{fp08}
H(\mathbf{r},\mathbf{r}';t)&\geq 0,\qquad \mathbf{r}\ne\mathbf{r}',\nonumber\\
\int\d^Dr\;H(\mathbf{r},\mathbf{r}';t)&=0.
\end{align}

Introducing the the basis vectors $|\mathbf{r}\rangle$ and their
dual basis covectors $\langle\mathbf{r}'|$,
\begin{align}\label{fp09}
\langle\mathbf{r}'|\mathbf{r}\rangle&=\delta(\mathbf{r}-\mathbf{r}'),\nonumber\\
\int\d^Dr\;|\mathbf{r}\rangle\langle\mathbf{r}|&=1,
\end{align}
one can write (\ref{fp01}) to (\ref{fp08}) in a more compact form.
To do so, one defines the vector $|\psi\rangle$ and the operators
$H$ and $U$ through
\begin{align}\label{fp10}
\psi(\mathbf{r})&=\langle\mathbf{r}|\psi\rangle,\nonumber\\
H(\mathbf{r},\mathbf{r}')&=\langle\mathbf{r}|H|\mathbf{r}'\rangle,\nonumber\\
U(\mathbf{r},\mathbf{r}')&=\langle\mathbf{r}|U|\mathbf{r}'\rangle,
\end{align}
and the covector $\langle\S|$ through
\begin{equation}\label{fp11}
\langle\S|:=\int\d^Dr\;\langle\mathbf{r}|.
\end{equation}
One would have then
\begin{align}\label{fp12}
\langle\mathbf{r}|\psi\rangle\geq&0,\\
\langle\S|\psi\rangle=&1,\\
|\psi(t)\rangle=&U(t,t')\,|\psi(t')\rangle,\\
\langle\mathbf{r}|U(t,t')|\mathbf{r}'\rangle\geq& 0,\\
\langle\S|U(t,t')=&\langle\S|,\\
U(t,t)=&1,\\
\frac{\d}{\d t}|\psi(t)\rangle=&H(t)\,|\psi(t)\rangle,\\
\frac{\partial U(t,t')}{\partial t}=&H(t)\,U(t,t'),\\
H(t):=&\left.\frac{\partial U(t,t')}{\partial t}\right|_{t=t'},\\
\langle\mathbf{r}|H(t)|\mathbf{r}'\rangle\geq& 0,\qquad \mathbf{r}\ne\mathbf{r}',\\
\langle\S|H(t)=&0.
\end{align}
These are in fact continuum analogs of a system with discrete
states, and can be readily obtained from that by suitably changing
summations to integrations.

The evolution is said to be local (in space), if the time
derivative of the density is expressible in terms of only a finite
number of space derivatives of the density. We restrict ourselves
to these cases. One then has
\begin{align}\label{fp23}
\frac{\partial\psi}{\partial t}=&\sum_{k=0}^n
(-1)^k\partial_{i_1}\cdots\partial_{i_k}
\left[f_{(k)}^{i_1\cdots i_k}\psi\right],\nonumber\\
=& \sum_{k=0}^n (-1)^k g_{(k)}^{i_1\cdots
i_k}\partial_{i_1}\cdots\partial_{i_k} \psi,
\end{align}
where $f^{i_1\cdots i_k}$'s, and $g^{i_1\cdots i_k}$'s are
functions related to each other. This is equivalent to
\begin{align}\label{fp24}
H(\mathbf{r},\mathbf{r}';t)=&\sum_{k=0}^n
(-1)^k\left[\partial_{i_1}\cdots\partial_{i_k}\delta(\mathbf{r}-\mathbf{r}')\right]\,
f_{(k)}^{i_1\cdots i_k}(\mathbf{r}'),\nonumber\\
=&\sum_{k=0}^n (-1)^k
\left[\partial_{i_1}\cdots\partial_{i_k}\delta(\mathbf{r}-\mathbf{r}')\right]\,
g_{(k)}^{i_1\cdots i_k}(\mathbf{r}).
\end{align}
The fact that $\psi$ is a probability density imposes constraints
on its evolution equation. If $\psi(\mathbf{r}_0,t_0)$ is zero,
then $\psi$ attains a minimum at $\mathbf{r}_0$ and its first
derivative vanishes at that point. As $\psi$ should remain
nonnegative everywhere, one should have
\begin{equation}\label{fp25}
\frac{\partial\psi(\mathbf{r}_0,t_0)}{\partial t}\geq 0.
\end{equation}
The initial condition $\psi(\mathbf{r},t_0)$ is arbitrary, except
that it should satisfy (\ref{fp01}). So the derivatives of order
more than two of $\psi$ with respect to $\mathbf{r}$ at $t=t_0$
and $\mathbf{r}=\mathbf{r}_0$ can attain arbitrary values. Suppose
that of the derivatives (of order more than two) which appear in
the right-hand side of (\ref{fp23}), only one is nonvanishing and
negative. If the absolute value of this derivative is large
enough, then the right-hand side of (\ref{fp23}) becomes negative,
which contradicts (\ref{fp25}). So, one concludes that only
derivatives of up to second order appear in the right-hand side of
(\ref{fp23}). This is discussed in \cite{Paw}. Moreover, at a
point where $\psi$ vanishes, the matrix of its second derivative
should be positive semi-definite, as at that point $\psi$ attains
its minimum. The matrix $g_{(2)}$ should be such that the
time-derivative of $\psi$ be nonnegative at such a point. This
means that $g_{(2)}$ should be positive semi-definite.

So, the eveolution equation for $\psi$ is
\begin{align}\label{fp26}
\frac{\partial\psi}{\partial t}=&f_{(0)}\,\psi-
\partial_i\left[f_{(1)}^i\,\psi\right]+\partial_i \partial_j\left[f_{(2)}^{ij}\,\psi
\right],\\ \label{fp27} \frac{\partial\psi}{\partial
t}=&g_{(0)}\,\psi-g_{(1)}^i\,\partial_i\psi+g_{(2)}^{ij}\,\partial_i
\partial_j\psi.
\end{align}
Integrating (\ref{fp26}) on the whole space with
$\psi(\mathbf{r},t)=\delta(\mathbf{r}-\mathbf{r}_0)$, and using
(\ref{fp01}), one deduces that $f_{(0)}(\mathbf{r}_0)$ vanishes.
As $\mathbf{r}_0$ is arbitrary, this means that
\begin{equation}\label{fp28}
f_{(0)}=0.
\end{equation}
So (\ref{fp26}) is simplified to
\begin{equation}\label{fp29}
\frac{\partial\psi}{\partial t}=-
\partial_i\left[f_{(1)}^i\,\psi\right]+\partial_i \partial_j\left[f_{(2)}^{ij}\,\psi
\right].
\end{equation}
Comparing this with (\ref{fp27}), one obtains the relations
between $f_{(k)}$'s and $g_{(k)}$'s:
\begin{align}\label{fp30}
0=&g_{(0)}+ \partial_i g_{(1)}^i+ \partial_i \partial_j g_{(2)}^{ij},\nonumber\\
f_{(1)}^i=&g_{(1)}^i+ 2\,\partial_j g_{(2)}^{ij},\nonumber\\
f_{(2)}^{ij}=&g_{(2)}^{ij}.
\end{align}

Introducing the operators $\mathbf{R}$ and $\mathbf{P}$ through
\begin{align}\label{fp31}
R^i|\mathbf{r}\rangle:=&r^i|\mathbf{r}\rangle,\nonumber\\
P_i|\mathbf{r}\rangle:=&\partial_i|\mathbf{r}\rangle,
\end{align}
it is seen that
\begin{align}\label{fp32}
[R^i,R^j]=&0,\nonumber\\
[P_i,P_j]=&0,\nonumber\\
[R^i,P_j]=&\delta^i_j,
\end{align}
and
\begin{equation}\label{fp33}
\langle\S|P_j=0,
\end{equation}
and the Hamiltonian for the system can be written as
\begin{align}\label{fp34}
H=&P_i\,f_{(1)}^i(\mathbf{R})+P_i\,P_j\,f_{(2)}^{ij}(\mathbf{R}),\nonumber\\
=&-\left[\partial_ig_{(1)}^i+\partial_i\partial_jg_{(2)}^{ij}\right](\mathbf{R})
+g_{(1)}^i(\mathbf{R})\,P_i+g_{(2)}^{ij}(\mathbf{R})\,P_i\,P_j,
\end{align}

\section{Random fields}
Generalization of the arguments of the previous section to the
case of fields is quite straight forward. A random field is a
collection of random variables defined at each point of the space.
One only needs to consider the following correspondences.
\begin{align}\label{fp35}
|\mathbf{r}\rangle\to&|\varrho\rangle,\nonumber\\
R^i\to&\rho(\mathbf{x}),\nonumber\\
P_i\to&\Pi(\mathbf{x}),\nonumber\\
\sum_i\to&\int\d^n x,\nonumber\\
\partial_i\to&\frac{\delta}{\delta\varrho(\mathbf{x})},\nonumber\\
\int\d^D r\to&\int D\varrho,
\end{align}
where $\rho$ is the random field operator, and one has
\begin{align}\label{fp36}
\rho(\mathbf{x})\,|\varrho\rangle=&\varrho(\mathbf{x})\,|\varrho\rangle,\nonumber\\
\Pi(\mathbf{x})\,|\varrho\rangle=&
\frac{\delta}{\delta\varrho(\mathbf{x})}|\varrho\rangle,\nonumber\\
[\rho(\mathbf{x}),\rho(\mathbf{y})]=&0,\nonumber\\
[\Pi(\mathbf{x}),\Pi(\mathbf{y})]=&0,\nonumber\\
[\rho(\mathbf{x}),\Pi(\mathbf{y})]=&\delta(\mathbf{x}-\mathbf{y}),\nonumber\\
\langle\varrho|\varrho'\rangle=&\delta(\varrho-\varrho'),\nonumber\\
\langle\S|:=&\int D\varrho\;\langle\varrho|,\nonumber\\
\langle\S|\,\Pi(\mathbf{x})=&0.
\end{align}
It then follows that (\ref{fp34}) is changed to
\begin{equation}\label{fp37}
H=\int\d^n x\;\Pi(\mathbf{x})\,f_{(1)}[\mathbf{x};\rho]+\int\d^n
x\, \d^n
y\;\Pi(\mathbf{x})\,\Pi(\mathbf{y})\,f_{(2)}[\mathbf{x},\mathbf{y};\rho],
\end{equation}
and $f_{(2)}$ is positive semi-definite, that is, one has for
arbitrary $h$ and $\varrho$,
\begin{equation}\label{fp38}
\int\d^n x\,\d^n y\;h(\mathbf{x})\,h(\mathbf{y})
\,f_{(2)}[\mathbf{x},\mathbf{y};\varrho]\geq 0.
\end{equation}
The evolution equation for $\psi$ (the probability distribution
functional for $\rho$) is given analogues to (\ref{fp26}):
\begin{align}\label{39}
\frac{\partial\psi}{\partial t}=&-\int\d^n x\;\frac{\delta}{\delta
\varrho(\mathbf{x})}\left(f_{(1)}[\mathbf{x};\varrho]\,\psi[\varrho]\right)
\nonumber\\
&+\int\d^n x\,\d^n
y\;\frac{\delta^2}{\delta\varrho(\mathbf{x})\,\delta\varrho(\mathbf{y})}
\left(f_{(2)}[\mathbf{x},\mathbf{y};\varrho]\,\psi[\varrho]\right).
\end{align}
One may also obtain the evolution equation for the expectation
value of the field
($\langle\rho(\mathbf{z})\rangle=\langle\S|\rho(\mathbf{x}|\psi\rangle$).
One has
\begin{align}\label{fp40}
\frac{\partial\rho(\mathbf{z})}{\partial t}=& [\rho(\mathbf{z}),H],\nonumber\\
=&f_{(1)}[\mathbf{z};\rho]+2\int\d^n
x\;\Pi(\mathbf{x})\,f_{(2)}[\mathbf{x},\mathbf{z};\rho],
\end{align}
from which
\begin{equation}\label{fp41}
\frac{\partial }{\partial t}\langle
\rho(\mathbf{z})\rangle=\langle f_{(1)}[\mathbf{z};\rho]\rangle.
\end{equation}
It is seen that $f_2$ has no explicit contribution in the
expectation value. It does, however, have implicit contribution
through $\rho$. A special case is $f_2=0$. It can be shown then
that if $f_{(1)}$ is a local functional of $\rho$, that is if
$f_{(1)}[\mathbf{z};\rho]$ depends on only $\rho$ and a finite
number of its derivatives in $\mathbf{z}$, then if the system is
initially uncorrelated in space,
\begin{equation}\label{fp42}
\langle\rho(\mathbf{x})\,\rho(\mathbf{y})\rangle=\langle\rho(\mathbf{x})\rangle\,
\langle\rho(\mathbf{y})\rangle,\qquad \mathbf{x}\ne\mathbf{y},
\end{equation}
then it remains uncorrelated in space. For a system initially
uncorrelated, and for $\mathbf{x}\ne\mathbf{y}$, one has initially
\begin{align}\label{fp43}
\frac{\partial}{\partial t}
\langle\rho(\mathbf{x})\,\rho(\mathbf{y})\rangle=&\langle
f_{(1)}[\mathbf{x};\rho]\,\rho(\mathbf{y})\rangle+
\langle\rho(\mathbf{x})\,f_{(1)}[\mathbf{y};\rho]\rangle,\nonumber\\
=&\langle
f_{(1)}[\mathbf{x};\rho]\rangle\,\langle\rho(\mathbf{y})\rangle+
\langle\rho(\mathbf{x})\rangle\,\langle f_{(1)}[\mathbf{y};\rho]\rangle,\nonumber\\
=&\left[\frac{\partial}{\partial t}\langle
\rho(\mathbf{x})\rangle\right]\,\langle\rho(\mathbf{y})\rangle+
\langle\rho(\mathbf{x})\rangle\,\left[\frac{\partial}{\partial
t}\langle\rho(\mathbf{x})\rangle\right],
\end{align}
which shows that initially
\begin{equation}\label{fp44}
\frac{\partial}{\partial
t}\left[\langle\rho(\mathbf{x})\,\rho(\mathbf{y})\rangle-
\langle\rho(\mathbf{x})\rangle\,\langle\rho(\mathbf{y})\rangle\right]=0.
\end{equation}
So the system remains uncorrelated.

Applying the approximation
\begin{equation}\label{fp45}
\langle f_{(1)}[\mathbf{x},\rho]\rangle\approx
f_{(1)}[\mathbf{x},\langle\rho\rangle],
\end{equation}
the evolution equation for the expectation value of $\rho$ becomes
\begin{equation}\label{fp46}
\frac{\partial }{\partial t}\langle\rho(\mathbf{x})\rangle=
f_{(1)}[\mathbf{x};\langle\rho\rangle].
\end{equation}
If $f_{(1)}$ is local, then the right-hand side of the above
equation is  the action of a differential operator on
$\langle\rho(\mathbf{x})\rangle$.
\section{Simple examples}
Consider a one-dimensional system for which $f_{(1)}$ is local,
apply the approximation (\ref{fp45}), and assume that
\begin{equation}\label{fp47}
f_{(1)}[x,\rho]=\frac{\partial^2\rho}{\partial x^2}+F[\rho(x)].
\end{equation}
Then the evolution equation for the expectation of $\rho(x)$ is
\begin{equation}\label{fp48}
\frac{\partial\langle\rho(x,t)\rangle}{\partial
t}=\frac{\partial^2\langle\rho(x,t)\rangle}{\partial
x^2}+F[\langle\rho(x,t)\rangle].
\end{equation}
Of special interest are the time-dependent solutions to this
equation:
\begin{equation}\label{fp49}
\frac{\d^2u(x)}{\d x^2}+F[u(x)]=0,
\end{equation}
where
\begin{equation}\label{fp50}
u(x):=\langle\rho(x)\rangle.
\end{equation}
Introducing the potential function $V$ through
\begin{equation}\label{fp51}
\frac{\d V}{\d u}:=F(u),
\end{equation}
one can once integrate (\ref{fp49}) to obtain
\begin{equation}\label{fp52}
\frac{1}{2}\,\left(\frac{\d u}{\d x}\right)^2+V(u)=\mathrm{const.}
\end{equation}
An obvious solution to this, or (\ref{fp49}), is
\begin{equation}\label{fp53}
u(x)=u_0,
\end{equation}
where $u_0$ is z zero of $F$. A kink solution is a solution which
tends to constant values at infinity, but the value at $+\infty$
is different from the value at $-\infty$:
\begin{equation}\label{fp54}
\lim_{x\to\pm\infty} u(x)=u_{\pm},\qquad u_+\ne u_-.
\end{equation}
For such a solution to exist, one should have
\begin{align}\label{fp55}
V(u_-)=&V(u_+),\nonumber\\
\left.\frac{\d V}{\d u}\right|_{u_\pm}=&0,
\end{align}
$V$ should be less than $V(u_-)=V(u_+)$ if $u$ is between $u_-$
and $u_+$. A simple example is
\begin{equation}\label{fp56}
V(u)=-\alpha\,(u-u_-)^2\,(u-u_+)^2,
\end{equation}
where $u_{\pm}$ and $\alpha$ are constants and $\alpha$ is
positive.

A multi-kink solution is a solution that tends to constant
(different) values at plus and minus infinity, and is constant (or
approximately constant) in some other intervals as well. Such
solutions exist if there are some points $u_i$ between $u_-$ and
$u_+$, where the potential $V$ becomes nearly equal to
$V(u_-)=V(u_+)$. One then has a solution which tends to constants
at infinity, and is nearly equal to $u_i$ and nearly constant for
a large region. An example is
\begin{equation}\label{fp57}
V(u)=-[\alpha+\beta\,(u-u_0)^2]\,(u-u_-)^2\,(u-u_+)^2,
\end{equation}
where $u_{\pm}$, $\alpha$, and $\beta$ are constants, $\alpha$ and
$\beta$ are positive, and $u_0$ is between $u_-$ and $u_+$. If
$\alpha$ is small, then there is an approximately double-kink
solution.

A special case is when $V(u_0)$ is exactly equal to $V(u_\pm)$. In
this case, if $V$ has second derivative in $u_0$, then the
double-kink solution disappears and two distinct single-kink
solutions emerge: one with between $u_-$ and $u_0$, the other
between $u_0$ and $u_+$. However, if the first derivative of $V$
near $u_0$ tends to zero like $|u-u_0|^\gamma$ with $\gamma<1$,
then there is a solution with a flat part at $u_0$ of a finite
width. An example is
\begin{equation}\label{fp58}
V(u)=-\beta\,|u-u_0|^{1+\gamma}\,(u-u_-)^2\,(u-u_+)^2,
\end{equation}
where $0<\gamma<1$.

These are stationary solutions. It is easy to see that a
stationary solution is readily converted to a travelling solution
using a Galilean transformation. This transformation only changes
the time-derivative $\frac{\partial}{\partial t}$ to
$\frac{\partial}{\partial t} +\mathbf{v}\cdot\nabla$, where
$\mathbf{v}$ is the corresponding velocity parameter, and is in
fact the velocity of the travelling solution.

\end{document}